\documentclass[reprint,aps,prd,twocolumn,groupedaddress,nofootinbib]{revtex4-2}

\usepackage{amsmath,amssymb,amsthm,graphicx}
\usepackage[dvipsnames]{xcolor}
\usepackage{multirow}
\usepackage{adjustbox}
\usepackage{array}
\usepackage{tabularx} 
\usepackage{float}
\usepackage[utf8]{inputenc}
\usepackage{gensymb}
\usepackage{dcolumn}% Align table columns on decimal point
\usepackage{bm}% bold math
\usepackage{subfig}
\usepackage{blkarray}
\usepackage{afterpage}
\usepackage{caption}
\usepackage{hyperref}

\begin{document}

%\preprint{}

\title{H Marks the Spot: \\ Searching for Exotic Production of Higgs + X to Map Out New Physics}

%Searching for BSM Higgs Production to Map Out New Physics

%\author{Nathaniel Craig$^{1}$}
\author{Seth Koren$^{1,2}$}
\email{EFI Oehme Fellow (sethk@uchicago.edu)}
\author{Umut \"{O}ktem$^{1,3}$}
\email{ucoktem@ucdavis.edu}
\affiliation{$^{1}$ Department of Physics, University of California, Santa Barbara, CA 93106, USA}
\affiliation{$^{2}$ Enrico Fermi Institute, University of Chicago, Chicago, IL 60637, USA}
\affiliation{$^{3}$ Department of Physics and Astronomy, University of California, Davis, CA 95616, USA}

%\date{\today}

\begin{abstract}
We propose leveraging our proficiency for detecting Higgs resonances by using the Higgs as a tagging object for new heavy physics. In particular, we argue that searches for exotic Higgs production from decays of color-singlet fields with electroweak charges could beat current searches at the LHC which look for their decays to vectors. As an example, we study the production and decay of vector-like leptons which admit Yukawa couplings with SM leptons. We find that bounds from Run 2 searches are consistent with anywhere from hundreds to many thousands of Higgses having been produced in their decays over the same period, depending on the representation. Dedicated searches for these signatures may thus be able to significantly improve our reach at the electroweak energy frontier.
\end{abstract}

%\keywords{}

\maketitle

\section{Introduction \label{sec:intro}}
With the discovery of the Higgs boson at the Large Hadron Collider (LHC) \cite{Aad_2012, Chatrchyan_2012}, we have completely explored the map of the basic ingredients of the Standard Model (SM)---as drawn up by particle physicists in the 1960s and 70s. Through the combined efforts of thousands of physicists over decades of work, we indeed found treasure buried each place we were told to look,
%---not to mention a couple unmarked surprises!--- 
and this has opened a new era in the quest to understand particle physics. But we may not rest on our laurels---there is far more yet unknown, lying somewhere off where there be dragons, perhaps already sketched out in one of the many maps our theoretical cartographers have dreamt up. 
Questions from the origins of neutrino masses to the microphysics of dark matter to the mechanism of baryogenesis tell us we must keep searching.
And as in any good mystery, our discovery of this newest clue has raised even more questions regarding the origins of the Higgs. In particular, in recent years the much-loved top-down frameworks tying the Higgs to the above questions, in addition to classic ideas about the Higgs itself, have seen growing empirical tension. But while experimental signatures have remained elusive so far, there have been an array of new strategies proposed for looking beyond the SM (BSM).

One such strategy is leveraging our most recent clue about the universe to try to find further new evidence, as any good detective knows. Such strategies have already been investigated and implemented in multiple frontiers. Precision measurements of Higgs couplings can reveal clues about extended scalar sectors, as called for in a wide variety of models from supersymmetric extensions to more general two-Higgs-doublet models and beyond (e.g. \cite{kanemura2013,cheung2013higgs,Carena:2014nza,Henning:2014gca}). Out in the sky, the dynamics of the Higgs field during inflation can imprint features to be seen in cosmological experiments (e.g. \cite{Chen:2016hrz,East:2016anr,Kumar:2017ecc,Hook:2019zxa}) and later its tachyonic potential around the origin may lead to electroweak baryogenesis and signatures imprinted in gravitational waves (e.g. \cite{Huber:2008hg,Caprini:2009yp,Espinosa:2010hh,no2011large}). In colliders we already have nontrivial constraints on dark sectors from the Higgs invisible branching ratio (e.g. \cite{eboli2000observing,bai2012measuring,Ghosh:2012ep,Belanger:2013kya}), and can probe even more sharply BSM Higgs decays into long-lived particles (e.g. \cite{Strassler_2007,Strassler_2008,Curtin:2013fra,Curtin:2015fna,Clarke:2015ala,Csaki:2015fba,Alipour-Fard:2018lsf,Cheung:2019qdr,Fuchs:2020cmm}). Needless to say, the use of Higgs physics to probe a menagerie of epochs in diverse ways is but one of the many benefits to Particle Physics that the discovery of the Higgs has provided. 

Here our focus is on the application of this strategy to our exploration of the highest energies in view. In particular we study the prospects of searching for on-shell production of new particles with SM gauge charges via their decays to on-shell Higgses at colliders. Given our newfound skill at looking for products of Higgs decays, this can give us a powerful probe of new particles which decay into Higgs + X--where X is some SM particle--so can be seen in both 2- and 3-body invariant mass spectra. In fact this strategy of bootstrapping our knowledge of particle physics using previous discoveries has paid off before. Since the discovery of the W and Z bosons, both new fundamental particles---the top quark \cite{D0:1995jca,Abe:1995hr} and the Higgs---have been discovered among channels using the decay into an on-shell W/Z boson. Against this backdrop, the suggestion to look for signatures of new heavy particles decaying into on-shell Higgses is a particularly natural one. In this manner we can hope to sequentially build the steps in our stairway to heavy-duty understanding of the universe---standing firmly on our most-recently-built stair allows us to work on the next.

BSM production of the Higgs has seen previous study mainly in the specific contexts of weak-scale supersymmetry (e.g. \cite{Baer:2012ts,Howe:2012xe,Ghosh:2012mc,Arbey:2012fa,Bharucha:2013epa,Han:2013kza,Papaefstathiou:2014oja,Ellwanger:2014hca,Zhang:2015cta,Chakraborti:2015mra,Choudhury:2016lku,Liu:2020ctf,Liu:2020muv}) and other two-Higgs-doublet models (e.g. \cite{Kling:2015uba,Dermisek:2015hue,Dermisek:2016via,Dermisek:2019heo,Dermisek:2019vkc,Dermisek:2020gbr}), often coming along with large MET or a high multiplicity of final states from cascade decays (see also \cite{Kribs:2009yh,Kribs:2010hp} for earlier proposals to discover the Higgs with BSM production modes). We note in particular the work of Yu \cite{Yu:2014mda} on the difficult task of disentangling SM and BSM production modes using information from differential distributions, which will be essential in maximizing the efficiency of realistic searches. In this light, our main message is that the Higgs can do \textit{even more} for us---there is a far wider variety of new physics models which searches for exotic Higgs production can usefully help us probe.

In particular we are motivated by the desire to search for the next-lightest field with SM gauge charges, independent of any specific UV model. As is well known, the SM effective field theory (SMEFT) provides a powerful tool to organize the possible effects of such heavy fields on the SM fields, by adding irrelevant operators constructed from the SM fields and suppressed by a high-energy cutoff in a series expansion. However, it is not necessarily the case that new fields will have masses high enough that their effects will be amenable to an EFT analysis at a collider. Indeed, for the class of new fields of interest (to be discussed momentarily), the regime where a SMEFT analysis would be justified lies outside the possible reach of the LHC. Which is just to say that a benefit of the energy frontier is that we may produce these particles on-shell, where poles in their amplitudes lead to greatly enhanced, resonant production. This allows a fantastic complementarity with the precision frontier, where model-independent bounds placed directly on SMEFT operators can be translated into constraints on new fields \cite{Ellis_2018}.

At a collider we are thus in need of a more general framework for the effects of new particles which knows something more about their identities. While one may despair that we now have an infinitude of possible representations of look for, we may sensibly still use the SMEFT as an organizing principle. Ref \cite{de_Blas_2018} constructed a dictionary of single-field extensions of the SM by non-trivial, non-chiral representations which contribute to the SMEFT at dimension six or less. This requirement restricts to fields whose gauge charges allow them to have some non-minimal coupling to SM fields---past that dictated solely by gauge invariance. As a result of this, the field is generally destabilized and picks up decays to SM particles (barring some further symmetry). This is the interesting class of such representations which we expect to see in SM final states at a collider, so is the appropriate place to focus our attention. Of course there are far more phenomenological possibilities when more than one new field is added, and there are many ways to get more exotic signatures at a collider---from missing energy to long-lived invisible decays or disappearing tracks. Heuristically, our considerations should hold in extensions where there is some separation of scales between the lightest new state with SM charges and the mass of further new states, where the couplings are generic, and where the new states don't have large branching ratios to SM-neutral fields.

This dictionary of possible single-field extensions so serves as a concrete, well-motivated `to do' list by which we can measure our progress on exploring the general potential extensions to the Standard Model. So how is it going thus far? While the Large Hadron Collider has allowed us to constrain new fields charged under the strong force often to many TeV, the energy frontier for new electroweak states which are color singlets is far less explored by comparison. As it stands, many fields which interact with the electroweak gauge bosons have lower mass limits below 1 TeV---sometimes by a factor of a few. 

In the coming decades we may be able to make direct progress in this direction with a relatively cheap and relatively compact muon collider, which has seen increased interest of late (e.g. \cite{Han:2020pif,Han:2020uid,Han:2020uak,Chiesa:2020awd,Capdevilla:2020qel,Costantini:2020stv,Yin:2020afe,Buttazzo:2020eyl,Capdevilla:2021rwo,Liu:2021jyc}). At high energies such a machine effectively becomes an electroweak gauge boson collider, as the SM vectors dominate the muon PDF, and so produce particles with electroweak charges copiously. Thankfully, there are experimental collaborations actively working to develop the technology needed (e.g. \cite{Bogomilov:2019kfj,Delahaye:2019omf,Long:2020wfp,Bartosik:2020xwr}). And this is not merely construction of a streetlight to look under; there are fantastic possibilities that exploration of the uncolored electroweak energy frontier can reveal. While the hierarchy problem has these past decades motivated focusing on our reach for discovering new colored particles, the LHC increasingly rebuffs the traditional models spurring such hopes. But important questions such as baryogenesis and neutrino masses could be resolved by electroweak-charged but color-singlet physics that may be accessible at the electroweak energy frontier. In recent years there has also been much work on models addressing the hierarchy problem without new colored particles, including those which still have other SM charges (e.g. \cite{Burdman:2006tz,Cai:2008au,Burdman:2014zta,Serra:2019omd}). For a recent pedagogical introduction to the hierarchy problem which discusses the tension of classic solutions with LHC data and the novel directions this has spurred, see Ref. \cite{Koren:2020biu}. %%come up with something better

In the meantime, we must leave no stone unturned in squeezing the best constraints we can out of the LHC, which means optimizing the limited bandwidth of triggers and analyses, which in turn means understanding the best possible searches for each new particle whose presence we wish to explore. It is with that philosophy in the front of our minds that we explore the usage of Higgs resonances as a powerful probe of new fields with electroweak charges. As a result of the Goldstone equivalence theorem, heavy such fields which decay appreciably to electroweak vectors necessarily have appreciable branching ratios to Higgses. So events where those new fields are produced at the LHC include decays to all of the on-shell electroweak bosons. But this theorem doesn't dictate that the electroweak bosons themselves have similar branching ratios, as decay is clearly not a high energy process in the rest frame. Thus, while the electroweak gauge bosons often decay to light quarks, leading to jets that may blend in with the messy QCD background, or to neutrinos visible solely as missing energy, decays of the Higgs boson are thankfully more often distinctive at the LHC. Indeed, the Higgs has already been `discovered' in many decay channels, which is as much a testament to the genius of experimentalists as it is to the inapplicability of the Goldstone equivalence theorem. 

This leads us to expect that searching for the decays of new electroweak states to Higgses may offer better reach than searches solely for the decays to vectors. To explore and evince this, we consider the production of (Section \ref{sec:vllatlhc}) and constraints on (Section \ref{sec:higgsprod}) new vector-like fermions with Yukawa couplings to the SM Higgs and leptons. We choose an electroweak singlet, doublet, and triplet to examine---the other possible representations come from the replacing the Higgs in the Yukawa interaction with the Higgs conjugate of opposite hypercharge and shifting the hypercharge of the new field by 1. We reinterpret the CMS 13 TeV multilepton search \cite{Sirunyan:2019bgz} to find the current best energy-frontier constraints on these fields. %Of course low-energy constraints on SMEFT operators induced by these heavy fields still apply, but cannot probe Yukawa couplings 

We find that at the lower mass bound placed on each species by the CMS multilepton search, the same integrated luminosity would have produced from hundreds to many thousands of events containing decays to Higgses, as the eager reader may find in Figures \ref{fig:singlethiggs}-\ref{fig:tripletconstraints}. This is far greater than the number of signal events needed for conclusive discovery of the Higgs itself, though of course one never detects all the events, and we do not attempt to simulate a new search ourselves. But this still evinces enormous complementarity with low-energy SMEFT constraints, which strongly depend on the Yukawa couplings of the new fields and their mixing with SM leptons. In contrast, the on-shell production of these new states will take place primarily through their gauge couplings via a Drell-Yan-like process, and our above invocation of the Goldstone equivalence theorem guarantees that their on-shell decays produce copious Higgses almost \textit{regardless} of the size of the Yukawa coupling. At small values the decays become displaced and just require a new search strategy, but at some point the decay length exceeds our detectors, so with the LHC we may hope to probe down to roughly $\lambda \sim \sqrt{\frac{64\pi \hbar c}{M_\text{new} L_\text{max}}} \sim 10^{-9}$.

Our exploration here thus points to the benefits of search strategies focused on BSM production of Higgses. 
We leave it to the expert experimentalists to design and optimize particular searches for such exotic Higgs production. As we approach Run 3 and gear up for the high-luminosity LHC (HL-LHC) \cite{Bruning:2015dfu} and a few more decades of fantastic and productive collider physics at the LHC, we hope this work will allow searches offering greater insight into electroweak physics above the electroweak scale, and of course into the physics of the Higgs boson.

\section{Vector-Like Leptons at the LHC} \label{sec:vllatlhc}

In this section we study extensions of the SM with an additional vector-like fermion in the representations $(1,1,-1)$, $(1,2,-\frac{1}{2})$, or $(1,3,0)$ of $SU(3)_c \times SU(2)_L \times U(1)_Y$. In each case, these gauge charges allow a Yukawa coupling to the Higgs and either the left-handed or right-handed Weyl electron, and we consider coupling it solely to the first generation. The gauge charges also allow Drell-Yan-like on-shell pair production of the new states at the LHC, which will be the most important production process.

\subsection{The Singlet}

We first consider a vector-like electroweak singlet denoted by $E$ \cite{de_Blas_2018} that transforms as $(1,1,-1)$, using $\bar E$ for the conjugate Weyl field. Such a field appears in many well-motivated extensions to the SM, for example as a Kaluza-Klein excitation of the right-handed electron from a 5d orbifold (e.g. \cite{Scherk:1979zr,Pomarol:1998sd}).

Before electroweak symmetry breaking, the allowed Yukawa interaction takes the form:
\begin{equation}
    \mathcal{L}_\text{Yuk} = \lambda_E \Bar{E} \phi^\dagger L + h.c.,
\end{equation}
%\begin{equation}
%    \mathcal{L}_\text{Yuk} = \lambda_E \frac{h}{\sqrt{2}}\Bar{E} e + \frac{v}{\sqrt{2}} \lambda_E \Bar{E} e + c.c.
%\end{equation}
and after EWSB we replace $\phi \rightarrow \left(0, (h + v)/\sqrt{2}\right)$ to work in unitary gauge. In addition, since $E, \bar E$ are vector-like we may write the mass term $M'_E \bar{\psi}_E\psi_E$, now in four-component notation for compactness, where $\psi_E$ is the corresponding Dirac field. The covariant derivatives give the field's gauge interactions:
\begin{equation}
    \mathcal{L}_\text{Gauge} = g_1c_w \Bar{E} \sigma^\mu A_\mu \Bar{E}^\dagger -g_1s_w \Bar{E} \sigma^\mu Z_\mu \Bar{E}^\dagger +  h.c.,
\end{equation}
with similar terms for $E$. The Yukawa interaction leads to mass mixing of $E, \bar E$ with the SM electron after EWSB. The mass matrix is given by 
% \begin{equation}
% \mathcal{M}_E = 
%   \begin{pmatrix} m'_E &   \frac{\lambda_{E}v}{\sqrt{2}}  \\
% 0 & M'_{E} \end{pmatrix}
% \end{equation}
\begin{equation}
\mathcal{M}_E = 
\begin{blockarray}{c c c}
\bar e & \bar E & \\
\begin{block}{(c c) c}
  m'_E & \frac{\lambda_{E}v}{\sqrt{2}} & \hspace{0.5em} e \\
  0 & M'_E & \hspace{0.5em} E \\
\end{block}
\end{blockarray}
\end{equation}
\noindent where we have labeled the rows and columns to make clear this is a Weyl mass matrix, and $m_E' = y_e' v/\sqrt{2}$ must be taken to have the value such that the light mass eigenstate has the observed electron mass $m_e$.

While in our numerical study we keep the full dependence on $\lambda_E$, we quote analytical expressions to lowest order in $\lambda_E v/M'_E$ for conceptual clarity and because $\lambda_E v/M'_E \ll 1$ in the allowed parameter space, as we'll see in Section \ref{sec:higgsprod}. The `bare electron mass' parameter $m_E'$ relates to the physical electron mass $m_e$= 0.511 $\text{ MeV}/c^2$ by
\begin{equation}
    m_e = m'_E \left( 1 - \frac{\lambda^2_E  v^2}{4M^{'2}_E} \right),
\end{equation}
and the larger mass eigenvalue $M_E$ is given by
\begin{equation}
    M_E = M'_E\left(1 + \frac{\lambda^2_E v^2}{4M^{'2}_E}\right),
\end{equation}
where we see the lighter eigenvalue has been shifted down and the heavier one has been shifted up. As for the mass eigenstates, we must find two unitary matrices $L$ and $R$, which rotate the left handed and right handed leptons respectively, that diagonalize $\mathcal{M}^2$ \cite{Haber,Haber:2020wco}. We first find $R$ by demanding $R^\dagger \mathcal{M}_E^\dagger \mathcal{M}_E R = M^2_D$, where $M_D$ is diagonal, and then we can find the left handed mixing using $L=(\mathcal{M}_E^\intercal)^{-1}R^* M_D$. Each mixing is here defined by a single angle, and at first order in $\lambda_E v/M'_E$ we have
\begin{equation}
    \theta_L = - \frac{\lambda_E v}{\sqrt{2}M'_E}, \qquad \theta_R = \frac{\lambda_E v}{\sqrt{2}M'_E} \frac{m_e}{M'_E},
\end{equation}
where the right handed mixing requires an extra mass insertion as the Yukawa interaction couples solely to the left handed electron. For compactness, we'll denote $\cos(\theta_L)\equiv c_L, \sin(\theta_R) \equiv s_R$, etc. We can then change variables to the mass eigenstates, which by abuse of notation we also refer to as $e$ and $E$ for the lighter and heavier eigenstate, letting the presence of trig functions of the mixing angles make clear we've rotated to the mass basis. At leading order this leads to mixed terms from the gauge interactions
\begin{equation}
 \mathcal{L}_\text{Gauge} \supset -\frac{g_2}{2c_w} s_L c_L e^\dagger \bar{\sigma}^\mu Z_\mu E -\frac{g_2}{\sqrt{2}} s_L  W^{+}_\mu \nu^\dagger \Bar{\sigma}^\mu E + h.c.
\end{equation}
The effects of the right handed components $\bar e, \bar E$ mixing is negligible,  since $\theta_R \ll \theta_L$, so we have left off their additive effects in the above expression. 

\begin{figure}[t]
        \centering
        \includegraphics[width = 8. cm]{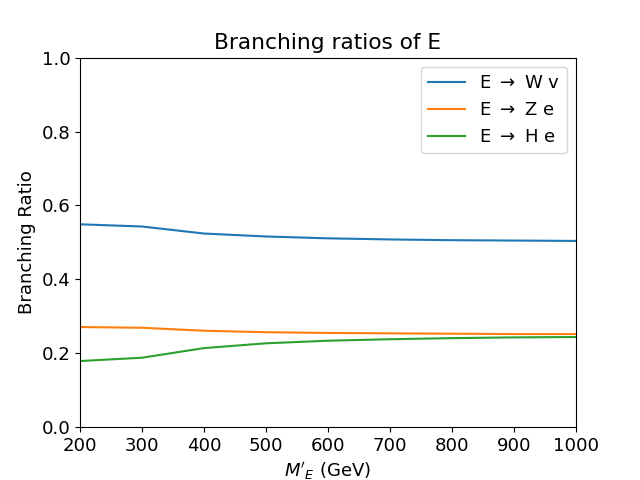}
  \caption{Branching ratios of the $SU(2)$ singlet $E$ as a function of the mass parameter $M'_E$.}
  \label{fig:singletBR}
\end{figure}

Together, these lead to the following expressions for the partial decay widths of the mass eigenstate $E$:
\begin{align}
    \Gamma(E \rightarrow Z e) &= \frac{ s^2_L c^2_L}{32 \pi}\frac{M^{'3}_E}{v^2}\left(1-\frac{M^2_Z}{M^{'2}_E}\right)^2\left(1+2\frac{M^2_Z}{M^{'2}_E}\right) \\
    \Gamma(E \rightarrow h e) &=  \frac{\lambda^2_{E}M'_E c^2_L c^2_R}{64 \pi}\left(1-\frac{M^2_H}{M^{'2}_E}\right)^2 \\
    \Gamma(E \rightarrow W \nu_e) &= \frac{ s^2_L}{16 \pi}\frac{M^{'3}_E}{v^2}\left(1-\frac{M^2_W}{M^{'2}_E}\right)^2\left(1+2\frac{M^2_W}{M^{'2}_E}\right)
\end{align}
The branching ratios are plotted in Figure \ref{fig:singletBR}, which can be seen to evince the Goldstone equivalence theorem for large $M'_E$. Note that in the regime where $\frac{\lambda_E v}{M'_E}\ll 1$ all the decay widths have the same dependence on $\lambda_E$, so the branching ratios will be independent of its value. 

\begin{figure}[t]
        \centering
        \includegraphics[width = 8. cm]{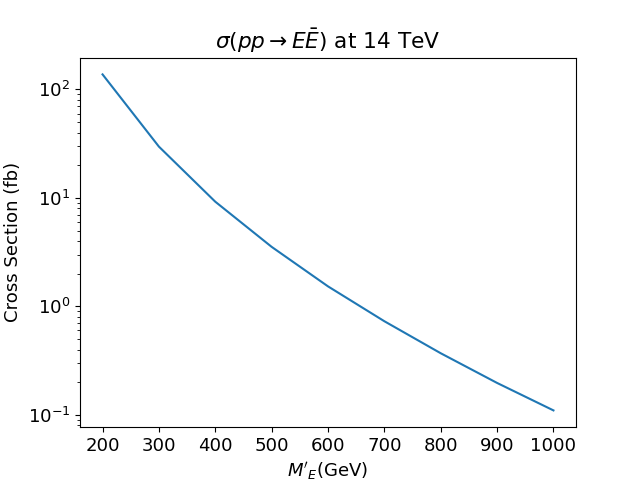}
  \caption{Production cross section of $E\bar E$ at the $14$ TeV LHC.}
  \label{fig:singletxsec}
\end{figure}

In this region of parameter space, the Drell-Yan-like double production through the gauge coupling is significantly larger than single production rates, so we restrict our attention to that mode. We note that this means our cross-section for on-shell production followed by decay into at least one Higgs is independent of the Yukawa coupling.
To calculate cross sections at the LHC, we implement the model in \texttt{FeynRules} \cite{Alloul_2014} as input to \texttt{MadGraph 5} \cite{Alwall_2011} for Monte Carlo simulation. We plot the production cross section as a function of $M'_E$ in Figure \ref{fig:singletxsec} and note that we are in agreement with \cite{Kumar_2015}.

In Section \ref{sec:higgsprod} we will study the constraints on this model from Run 2 of the LHC and compare the prospects for BSM Higgs production, but for now we will proceed to discuss the next representation.

\subsection{The Doublet}
Next on the list, we consider a vector-like SU(2) doublet $\Delta_1$, transforming as $(1,2,-\frac{1}{2})$, with its partner $\bar\Delta_1$ \cite{de_Blas_2018}. After EWSB we'll find we have one neutral field and one charged field, so we denote the corresponding left-handed Weyl fields:
 \begin{align}
    \Delta_1 &= \begin{bmatrix}
           N \\
           E
         \end{bmatrix},
  \end{align}
and similarly for $\bar \Delta_1$. Since we have a vector-like pair, a mass term is allowed in the Lagrangian. 
\par
Such doublets are required in certain grand unification schemes (e.g. $E_6$ \cite{Gursey:1975ki,Gursey:1978fu}) or may appear simply as the supersymmetric partners of the two Higgs doublets in the MSSM, with $\Delta_1 \simeq \tilde{H}_u, \bar \Delta_1 \simeq \tilde{H}_d$. 

Before electroweak symmetry breaking, the allowed Yukawa interaction can be written as:
\begin{equation}
    \mathcal{L}_\text{Yuk} = -\lambda_{\Delta_1} \bar \Delta_1 \phi \bar{e} + h.c.
\end{equation}
 \begin{figure}[t]
        \centering
        \includegraphics[width = 8. cm]{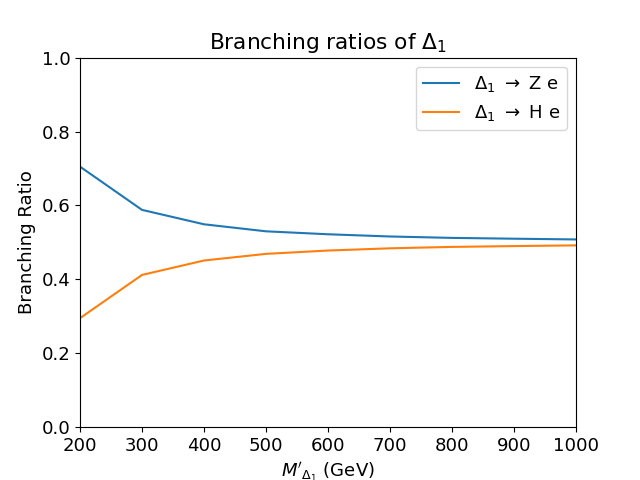}
            \caption{Branching ratios of the charged component of the $SU(2)$ doublet $\Delta_1$ as a function of the mass parameter $M'_{\Delta_1}$.}
    \label{fig:delbr}
\end{figure}
Unlike the previous case the coupling is now to the right-handed electron. This interaction leads to mixing solely for the charged component $E$. \par
The gauge interaction terms are now:
\begin{align}
    \mathcal{L}_\text{Gauge} &= \frac{g_2}{\sqrt{2}}W^{+}_\mu N^\dagger\bar{\sigma}^\mu E \nonumber\\
    &+\frac{g_2}{c_w}Z_\mu \left(-\frac{1}{2}+s^2_w\right) E^\dagger\bar{\sigma}^\mu E \nonumber\\
    &+ \frac{g_2}{2 c_w}Z_\mu N^\dagger\bar{\sigma}^\mu N \nonumber\\
    &+ e A_\mu E^\dagger\bar{\sigma}^\mu E,
\end{align}

\noindent where there are additional similar terms for the right handed fields $\bar E, \bar N$.  The mass matrix in this case is given by:
\begin{equation}
\mathcal{M}_{\Delta} = 
  \begin{pmatrix} m'_E &   0  \\
\frac{\lambda_{\Delta_1}v}{\sqrt{2}} & M'_{\Delta_1} \end{pmatrix}
\end{equation}

As before, the parameter $m'_E$ must be chosen to ensure the lighter eigenvalue of $\mathcal{M}_{\Delta}^2$ corresponds to the physical electron mass $m_e$. We repeat the same procedure as above and find the eigenvalues similarly:
\begin{equation}
    m_e = m'_E \left(1 - \frac{\lambda_E^2 v^2 }{4M^{'2}_{\Delta_1}} \right), \quad M_{\Delta_1} = M'_{\Delta_1}\left(1+\frac{\lambda^2_{\Delta_1}v^2}{4M^{'2}_{\Delta_1}}\right).
\end{equation}

As for the mass eigenstates, this time the mixing angles to first order in $\lambda_{\Delta_1}$ are: \par
\begin{equation}
     \theta_R = \frac{\lambda_{\Delta_1}v}{\sqrt{2}M'_{\Delta_1}}, \qquad \theta_L = -\frac{\lambda_{\Delta_1}v}{\sqrt{2}M'_{\Delta_1}} \frac{m_e}{M'_{\Delta_1}},
\end{equation}
with now the left-handed mixing being suppressed by a mass insertion. Using the previous convention for compactifying the mixing angle expressions, we expand the gauge interaction terms in terms of mass eigenstates to first order, and we find the mixed gauge interaction terms in the mass basis:
\begin{equation}
 \mathcal{L}_\text{Gauge} \supset -\frac{g_2}{2c_w} s_R c_R  \bar{E}^\dagger \Bar{\sigma}^\mu Z_\mu \bar{e} - \frac{g_2}{\sqrt{2}}s_R W^{+}_\mu \bar{N} \sigma^\mu \bar{e}^\dagger + h.c.,
\end{equation}
\noindent with the photon vertex cancelling out. 

\begin{figure}[t]
        \centering
        \includegraphics[width = 8. cm]{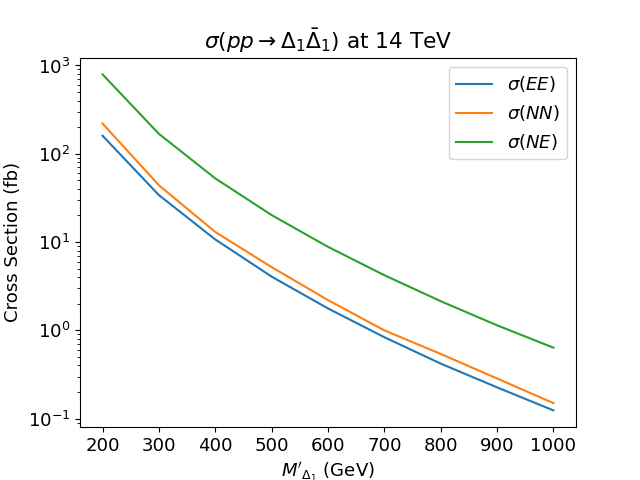}
  \caption{Production cross section of the components of $\Delta_1 \bar \Delta_1$ at the 14 TeV LHC.}
  \label{fig:delxsec}
\end{figure}

We can now calculate decay widths for the mass eigenstates $E$, $N$, which we find to be:
\begin{equation}
    \Gamma(E \rightarrow Z e) = \frac{s^2_R c^2_R}{32 \pi}\frac{M^{'3}_{\Delta_1}}{v^2} \left(1-\frac{M^2_Z}{M^{'2}_{\Delta_1}} \right)^2 \left(1+2\frac{M^2_Z}{M^{'2}_{\Delta_1}} \right)
\end{equation}

\begin{equation}
    \Gamma(E \rightarrow h e) =  \frac{\lambda^2_{\Delta_1}M'_{\Delta_1} c^2_L c^2_R}{64 \pi} \left(1-\frac{M^2_H}{M^{'2}_{\Delta_1}} \right)^2
\end{equation}

\begin{equation}
    \Gamma(N \rightarrow W e) = \frac{ s^2_R}{16 \pi}\frac{M^{'3}_{\Delta_1}}{v^2} \left(1-\frac{M^2_W}{M^{'2}_{\Delta_1}} \right)^2 \left(1+2\frac{M^2_W}{M^{'2}_{\Delta_1}}\right)
\end{equation}
We plot the branching ratios of the charged component as a function of $M'_{\Delta_1}$ in Figure \ref{fig:delbr}.

We simulate the $pp \rightarrow \Delta_1 \bar{\Delta}_1$ cross section in the same manner as above, and find the production cross sections in Figure \ref{fig:delxsec}, which are again in agreement with \cite{Kumar_2015}. We note that $N$ decays solely to $W e$ so its production does not lead to anomalous Higgs events.

We turn now to the final representation we study. \par

\subsection{The Triplet}
Finally, we consider a vector-like SU(2) triplet fermion with zero hypercharge, which we'll denote by $\Sigma$ \cite{de_Blas_2018}. Such a field is familiar from, for example, type III Seesaw models \cite{Foot:1988aq}, 
where neutrino mass originates through mixing with $\Sigma$. 

We'll find it convenient to regard the adjoint representation as the traceless bifundamental, so we represent $\Sigma$ as a $2 \times 2$ matrix according to $\Sigma_{\alpha \bar \alpha} = \Sigma^a \sigma^a_{\alpha \bar \alpha}$, which we break into components as 
\begin{equation}
\Sigma = 
  \begin{pmatrix} 
  N/\sqrt{2} &  E^{+}  \\  
  E^{-} & -N/ \sqrt{2} 
  \end{pmatrix}
\end{equation}
The charged components may pair up to form a Dirac spinor, $\psi_E = \left(E^-, (E^+)^\dagger\right)^\intercal$, while we can write a Majorana spinor for the neutral component $\psi_N = \left(N, (N)^\dagger\right)^\intercal$.

\begin{figure}[t]
        \centering
        \includegraphics[width = 8. cm]{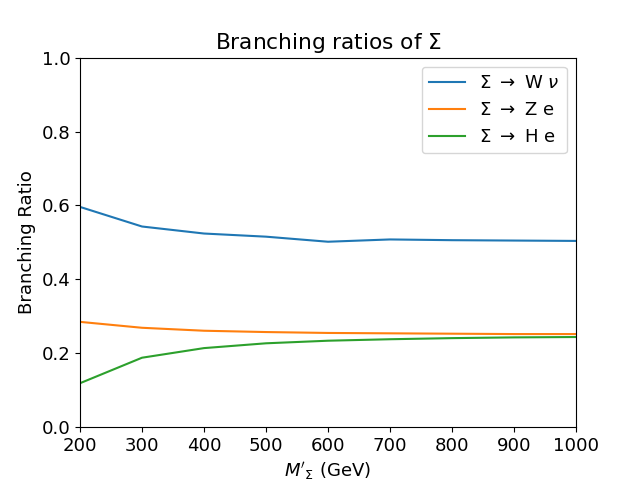}
            \caption{Branching ratios of the charged component of the $SU(2)$ triplet $\Sigma$ as a function of the mass parameter $M'_\Sigma$.}
    \label{fig:sigmabr}
\end{figure}

The Yukawa interaction now looks like 
\begin{equation}
    \mathcal{L}_\text{Yuk} = -\lambda_{\Sigma} \Tilde{\phi}^\dagger \Sigma  L + h.c.
\end{equation}
After EWSB this includes Yukawa interactions with both the left-handed electron and neutrino
\begin{equation}
    \mathcal{L}_\text{Yuk} = - \frac{ \lambda_{\Sigma} h N \nu_e}{2}  +  \frac{\lambda_{\Sigma}  h E^+ e}{\sqrt{2}} \ + \ h.c.
\end{equation}
Expanding the covariant derivative, we find the gauge interaction terms:
\begin{align}
    \mathcal{L}_\text{Gauge} &= g_2 c_w E^{+}\sigma^\mu Z_{\mu} (E^+)^{\dagger} + g_2 c_w (E^{-})^\dagger\bar{\sigma}^{\mu}Z_{\mu}E^{-}  \nonumber\\
    &+ e E^{+}\sigma^\mu A_{\mu} (E^+)^{\dagger} + e (E^{-})^\dagger\bar{\sigma}^{\mu}A_{\mu}E^{-} \\
    &-g_2  N \sigma^\mu W^{+}_{\mu}(E^+)^{\dagger} - g_2   N^{\dagger} \sigma^{\mu}W^{+}_{\mu}E^{-} + h.c.\nonumber
\end{align}
\par
This Yukawa interaction induces mixing of both the charged and neutral components of $\Sigma$ with SM leptons, so we now have two non-trivial mass matrices.

The charged mass matrix is given by:
\begin{equation}
\mathcal{M} = 
  \begin{pmatrix} m'_E &   \frac{\lambda_{\Sigma}v}{\sqrt{2}}  \\
0 & M'_{\Sigma} \end{pmatrix},
\end{equation}
\noindent and the Majorana mass matrix for the neutral states is:
\begin{equation}
\mathcal{M}_{0} = 
  \begin{pmatrix} m'_\nu &   \frac{\lambda_{\Sigma}v}{2}  \\
\frac{\lambda_{\Sigma}v}{2}  & M'_{\Sigma} \end{pmatrix}.
\end{equation}
\par
As we do not wish to specialize to type III seesaw models, we have included also a bare mass for the electron neutrino, $m'_\nu$, so that we can study general mixing with $\Sigma$ while keeping the neutrino mass fixed, which we set to $50$ meV.

\begin{figure}[t]
        \centering
        \includegraphics[width = 8. cm]{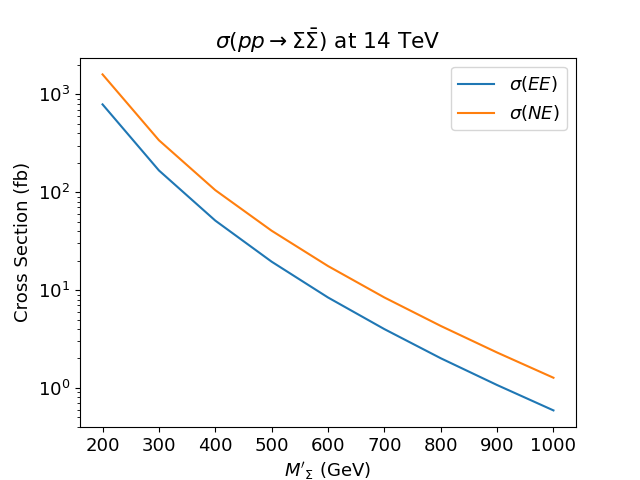}
  \caption{Production cross section of the components of $\Sigma \bar \Sigma$ at the $14$ TeV LHC.}
  \label{fig:sigmaprod}
\end{figure}

As before, the lighter mass eigenstates, consisting primarily of the SM leptons, have their masses shifted down from the Lagrangian parameters
\begin{align}
    m_e &= m'_E\left(1 - \frac{\lambda^2_\Sigma v^2}{4M^{'2}_\Sigma}\right) \\
    m_\nu &= m'_{\nu} - \frac{\lambda^2_\Sigma v^2}{4 M'_\Sigma},
\end{align}
and the heavier mass eigenvalues are likewise shifted up
\begin{align}
    M_{\Sigma^{\pm}} &= M'_\Sigma\left(1 +\frac{\lambda^2_{\Sigma}v^2}{4 M^{'2}_\Sigma}\right) \\
    M_{\Sigma^0} &= M'_\Sigma\left(1 +\frac{\lambda^2_{\Sigma}v^2}{4M^{'2}_\Sigma}\right),
\end{align}
where we note that this tree-level degeneracy will be broken by radiative corrections.

To diagonalize the charged mass matrix, we can repeat our procedure for the previous two cases and find
\begin{equation}
     \theta_L = -\frac{\lambda_\Sigma v}{ \sqrt{2}M'_\Sigma}, \qquad \theta_R = \frac{\lambda_\Sigma v}{\sqrt{2}M'_\Sigma} \frac{m_e}{M'_\Sigma},
\end{equation}
and for the neutral mass matrix we only have a single mixing
\begin{equation}
    \theta_N =  \frac{\lambda_\Sigma v}{2 M'_\Sigma}.
\end{equation}
\par
Shifting now to the mass basis with the same conventions as above, we find mixed gauge interactions
\begin{align}
    \mathcal{L}_\text{Gauge} &\supset  g_2  \left(c_N s_L-\frac{1}{\sqrt{2}} c_R s_N\right)N^{ \dagger} W^+_\mu\bar{\sigma}^\mu e  \nonumber\\ 
    &- g_2 s_N c_R (E^+)^\dagger W^+_\mu \sigma^\mu \nu   \nonumber\\
    &+ \frac{g_2}{2 c_w} s_N c_N N^{ \dagger} Z_\mu \sigma^\mu \nu  \nonumber\\
    &- \frac{g_2}{2 c_w} s_L c_R (E^-)^\dagger Z_\mu \sigma^\mu e.
\end{align}

\par
From this we can simply compute the partial decays widths, which for the neutral component $N$ are:
\begin{align}
    \Gamma(N \rightarrow Z \nu) &= \frac{s^2_N c^2_N}{32 \pi}\frac{M^{'3}_{\Sigma}}{v^2} \left(1-\frac{M^2_Z}{M^{'2}_{\Sigma}} \right)^2 \left(1+2\frac{M^2_Z}{M^{'2}_{\Sigma}} \right) \\
    \Gamma(N \rightarrow h \nu) &=  \frac{\lambda^2_{\Sigma}M'_\Sigma c^4_N}{128 \pi}\left(1-\frac{M^2_H}{M^{'2}_\Sigma} \right)^2 \\
    \Gamma(N \rightarrow W e) &= \frac{(2c_N s_L -\sqrt{2}c_R s_N)^2}{32 \pi} \times \\
    & \frac{M^{'3}_\Sigma}{v^2} \left(1-\frac{M^2_W}{M^{'2}_\Sigma}\right)^2 \left(1+2\frac{M^2_W}{M^{'2}_\Sigma}\right). \nonumber
\end{align}

And for the charged components $E^\pm$, the decay widths are:
\begin{align}
    \Gamma(E \rightarrow Z e) &= \frac{s^2_L c^2_L}{32 \pi}\frac{M^{'3}_{\Sigma}}{v^2} \left(1-\frac{M^2_Z}{M^{'2}_{\Sigma}} \right)^2 \left(1+2\frac{M^2_Z}{M^{'2}_{\Sigma}} \right) \\
    \Gamma(E \rightarrow h e) &=  \frac{\lambda^2_{\Sigma}M'_\Sigma c^2_L c^2_R }{64 \pi} \left(1-\frac{M^2_H}{M^{'2}_\Sigma} \right)^2\\
    \Gamma(E \rightarrow W \nu) &= \frac{s^2_N c^2_R }{8 \pi}
    \frac{M^{'3}_\Sigma}{v^2} \times \\ & \left(1-\frac{M^2_W}{M^{'2}_\Sigma} \right)^2 \left(1+2\frac{M^2_W}{M^{'2}_\Sigma} \right). \nonumber
\end{align}

We plot these branching ratios in Figure \ref{fig:sigmabr} and the double production cross section in Figure \ref{fig:sigmaprod}, finding agreement with \cite{Franceschini_2008}. With the stage now set, we turn to considering the constraints on these fields.

\section{Constraints and Discovery Potential} \label{sec:higgsprod}

As mentioned above, low-energy constraints on SMEFT operators can be straightforwardly used to constrain the presence of new BSM states. For many single field extensions the constraints have been systematically studied and Ref. \cite{Ellis_2018} gives bounds in the $(\lambda_i,M'_i)$ parameter space for our three representations which we reproduce in Table \ref{tab:table1}.
\begin{table}[h!]
  \begin{center}
    \caption{Constraints from the SMEFT Analysis \cite{Ellis_2018}.}
    \label{tab:table1}\large
   \begin{tabular}{|c|c|c|}
       \textbf{E} & \textbf{ $\Delta_1$} & \textbf{$\Sigma$}
      \\
        \hline
        $\frac{\lambda^2_E}{M^{'2}_E} \leq \frac{2.1\times 10^{-2}}{1 \text{ TeV}^2} $ &  $\frac{\lambda^2_{\Delta_1}}{M^{'2}_{\Delta_1}} \leq \frac{1.7\times 10^{-2}}{1 \text{ TeV}^2} $  & $\frac{\lambda^2_\Sigma}{M^{'2}_\Sigma} \leq \frac{4.5\times 10^{-2}}{1 \text{ TeV}^2} $  \\

       \end{tabular}
  \end{center}
\end{table}
\par

At the energy frontier, the best constraints on these models come from multilepton searches. Such searches, which require 3 or more final state leptons, essentially look for the decay modes to $W$ or $Z$ and their leptonic decays. There are thus large penalties from the high branching ratios to jets or neutrinos. The strongest constraint comes from the 13 \text{TeV} CMS multilepton search \cite{Sirunyan:2019bgz} taking advantage of the full Run 2 data set. The CMS analysis interprets the search in the context of a type III seesaw model with flavor-democratic couplings, and we use this model to validate our reinterpretation procedure. Details on the procedure and the validation are given in the appendix. For completeness we mention the ATLAS Run 2 dilepton + jets search  \cite{Aad:2020fzq}, which interprets their search with the same signal model and finds a bound which is nearly as strong, but we below reinterpret solely the leading bound from CMS. For earlier work related to bounds on the production of general vector-like leptons at colliders, see e.g. \cite{Buskulic:1996zs,Ackerstaff:1998si,Abreu:1998jw,Acciarri:1999bw,Achard:2001qw,CMS:2012ra,Aad:2015cxa,Aad:2015dha} and \cite{Thomas:1998wy,Franceschini_2008,del_Aguila_2008,Carpenter:2010sm,Carpenter:2010bs,Rajaraman:2010ua,Biggio:2011ja,Falkowski:2013jya,Altmannshofer:2013zba,Ma:2014zda,Dermisek:2014qca,Halverson:2014nwa,Kumar:2015tna,Chen:2016lsr,Bhattiprolu:2019vdu,Freitas:2020ttd,Bissmann:2020lge}

\afterpage{\clearpage}

\par
After Monto Carlo simulation from \texttt{MadGraph 5} we put the events through \texttt{Pythia} \cite{Sj_strand_2020} for decays and showering and then \texttt{Delphes} \cite{de_Favereau_2014} for detector simulation. Then we pass the resulting .lhco files through our reinterpretation of the CMS analysis using \texttt{MadAnalysis} \cite{Conte_2013} and translate our results into bounds on the mass parameter. We derive the 95\% confidence lower bounds for each field that are shown in Table \ref{tab:table2}.

\begin{figure}[p]
        \includegraphics[width = 7.65 cm]{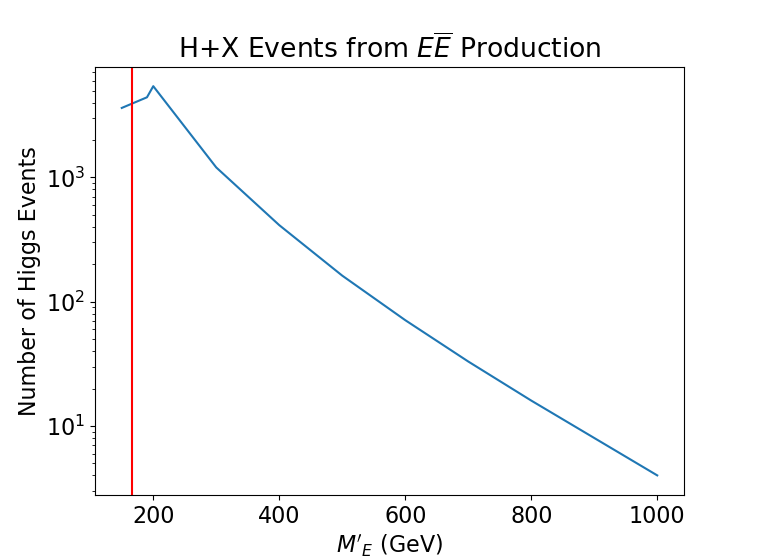}
            \caption{Number of anomalous Higgs events from production of the $SU(2)$ Singlet at Run 2 of the LHC. The red line denotes the lower mass bound placed by our reinterpretation of the CMS multilepton search.}
    \label{fig:singlethiggs}
\end{figure}
    
    \begin{figure}[p]
        \includegraphics[width = 7.65 cm]{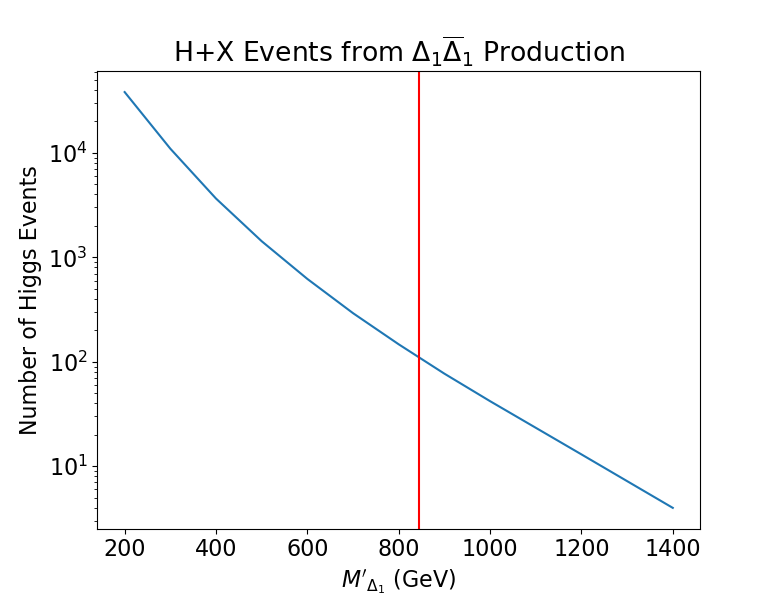}
            \caption{The same as Figure \ref{fig:singlethiggs} for the doublet.}
    \label{fig:doublethiggs}
    \end{figure}
    
    \begin{figure}[p]
        \includegraphics[width = 7.65 cm]{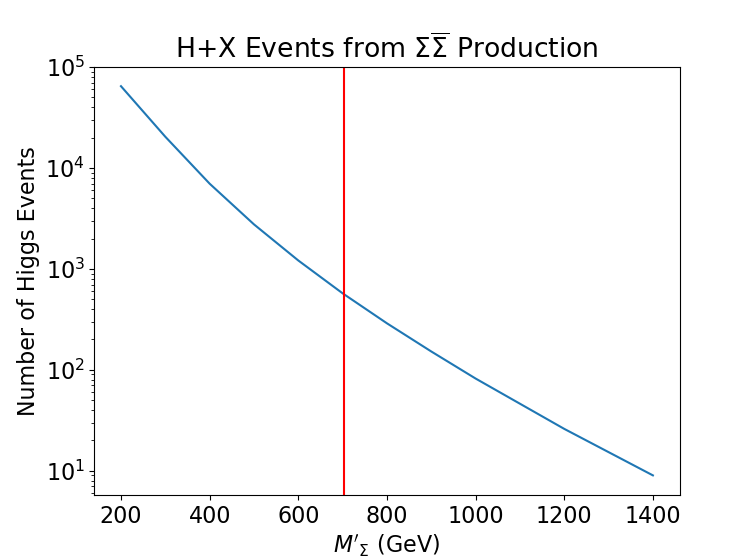}
            \caption{The same as Figure \ref{fig:singlethiggs} for the triplet.}
    \label{fig:triplethiggs}
    \end{figure}

\begin{figure}[p]
        \includegraphics[width = 7.65 cm]{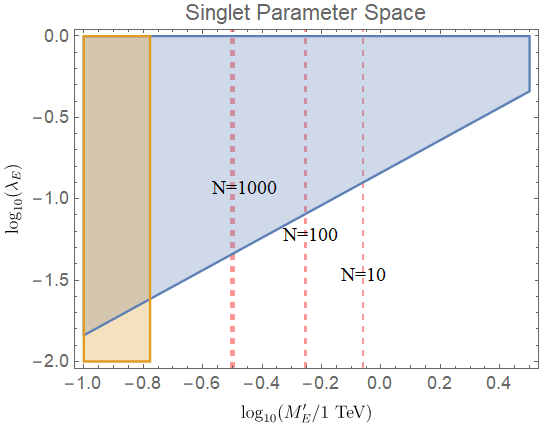}
            \caption{Bounds on the singlet parameter space from the SMEFT analysis \cite{Ellis_2018} (blue) and the CMS multilepton search \cite{Sirunyan:2019bgz} (orange). The dashed lines show the masses at which the labeled number of anomalous Higgs events would have been produced in Run 2. }
    \label{fig:singletconstraints}
    \end{figure}

\begin{figure}[p]
        \includegraphics[width = 7.65 cm]{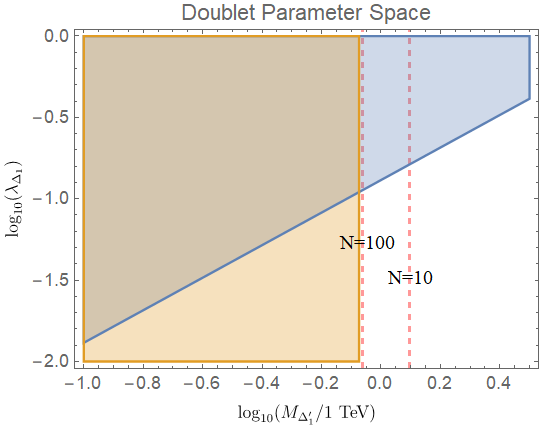}
            \caption{The same as Figure \ref{fig:singletconstraints} for the doublet.}
    \label{fig:doubletconstraints}
    \end{figure}

\begin{figure}[p]
        \includegraphics[width = 7.65 cm]{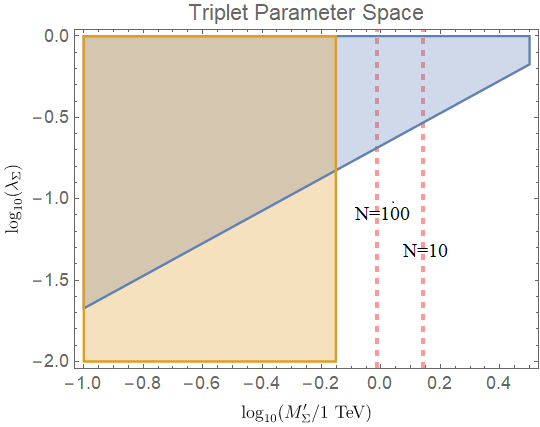}
            \caption{The same as Figure \ref{fig:singletconstraints} for the triplet.}
    \label{fig:tripletconstraints}
    \end{figure}

As discussed in Section \ref{sec:intro}, our newfound adeptness at identifying Higgs decays makes the Higgs itself a promising tagging object. To explore this, we study the pair production of our heavy leptons which include at least one decay to Higgs + X. While we do not attempt to mock up a realistic search, we will show that there is ample parameter space which the multilepton searches are unable to probe, yet which lead to a large rate of exotic Higgs production events. In Figures \ref{fig:singlethiggs}, \ref{fig:doublethiggs}, \ref{fig:triplethiggs} we plot the number of anomalous Higgs events each model would have produced in Run 2 of the LHC as a function of the mass of the new lepton. In Figures \ref{fig:singletconstraints}, \ref{fig:doubletconstraints}, \ref{fig:tripletconstraints} we show the constraints on these three models from the SMEFT analysis and our reinterpretation of the CMS multilepton search.  

\begin{table}[h]
  \begin{center}
    \caption{Bounds from the 13 TeV CMS Multilepton Search \label{tab:reintbounds}}
    \label{tab:table2}\large
    \begin{tabular}{|c|c|c|}
       \textbf{E} & \textbf{ $\Delta_1$} & \textbf{$\Sigma$}
      \\
        \hline
      $M'_E > 167 \text{ GeV}$ & $M'_{\Delta_1} > 844 \text{ GeV}$ &  $M'_\Sigma > 705 \text{ GeV}$  \\

       \end{tabular}
  \end{center}
\end{table}

As can be seen for the singlet from Figure \ref{fig:singlethiggs}, there are an abundance of exotic Higgs production events beyond where the Run 2 leptonic searches impose a bound. Figure \ref{fig:singletconstraints} illustrates for the same model how BSM Higgs production compares to bounds from SMEFT and the CMS multilepton search.
    
We note again that in calculating the numbers of Higgs
produced we use the branching ratios at $\lambda v/M'_E \ll 1$, which are independent of the value of $\lambda$. This is satisfied everywhere in the parameter space which is not already ruled out by SMEFT constraints. Resultingly we do not worry about the precise shape of the constraint from multilepton searches in the $\lambda v/M'_E \sim 1$ regime, which would solely modify the upper left corner of Figure \ref{fig:singletconstraints}.

The analogous results for the doublet and triplet may be found in Figures \ref{fig:doublethiggs}, \ref{fig:doubletconstraints} and \ref{fig:triplethiggs}, \ref{fig:tripletconstraints}, respectively. As evinced by these plots, there is a significant number of anomalous Higgs events produced outside of the region in parameter space ruled out by SMEFT bounds and current LHC searches. The benefit of a search for anomalous Higgs decays is most clear in the case of the singlet, but if a search can be designed with few background events, it should be able to probe untouched parameter space in all three models. The presence of resonances in 2- and 3-body invariant mass spectra should help make such a possibility practicable.

\section{Conclusion \label{sec:conc}}

The LHC is likely to be our clearest window into the energy frontier of the electroweak sector for at least the next couple decades. Obviously, we must wring every bit of constraining power out of the data we collect at the experiments. While the data can be fully analyzed at our leisure over the years when future machines are being built, the data collection itself confers urgency due to our finite capacity for processing and storing events. Over the next years, the LHC will deliver what is likely the last 14 TeV proton-proton collisions that humanity has access to for hundreds of years, if not far longer. Each event in every inverse femtobarn of integrated luminosity which is \textit{not} written to tape is an event which is lost forever to time, so we had better be damn sure we're keeping the right ones. Note particularly that a 100 TeV collider this century will not necessarily be able to make up this lost opportunity---data at different center-of-mass energies are complementary, and indeed data from the Tevatron has still been useful in the LHC era. So understanding the optimal signatures and triggers and searches to use to look for new physics is absolutely critical, as well as pressing.  

While the wide range of possible models of the universe makes systematically searching for new physics a cumbersome task, we have focused on a well-motivated restricted catalog of new fields to look for at a collider. We have argued that searches for signatures from the decays of on-shell new physics into Higgses can have greater reach than those currently performed toward the goal of probing new particles with SM charges. This can be motivated by quite general arguments for new massive color singlets with electroweak charges, for which the numbers of decays to Higgses and gauge bosons are similar. We have then evinced the opportunities available for such searches by reinterpreting current searches for the best bounds on a few representative new species, and showing that despite current constraints we may be allowing an enormous number of exotic Higgs production events to go unnoticed. This indicates that implementing a search for these Higgs + X final states may be able to improve the existing bounds on these models. Explicitly constructing such a search strategy we leave to the experimentalists. 

Of course there remains much more to be explored. We have not gone through each possibility in the catalog of interesting single-field extensions, particularly in restricting our attention to fermions in vector-like representations which couple to first-generation leptons.  Furthermore, our strategy of focusing on single-field extensions is convenient in providing a simple, manageable catalog, but the extension of the SM chosen by Nature may well have multiple new particles within reach of the energy frontier. One could imagine a wider array of phenomenologies when there is not a large gap between the lightest new state and the second-lightest. Mapping out the possibilities with two new states may be useful for pointing out in which regions of parameter space there are other signatures, not appearing in single-field extensions, that become the best way to constrain them. This may be a collider search for different final states, or displaced decays appearing when the states are nearly degenerate, or signals appearing in tabletop experiments or cosmological observations or elsewhere. 

The other clear direction of generalization is to study the production of and potential searches for species whose decays include other final states along with Higgses. While our focus has been on the electroweak energy frontier, analogous benefits may come from looking for decays of new colored fields to Higgses and jets, as has already been usefully done for e.g. vector-like quarks \cite{Sirunyan:2017usq,Aaboud:2018pii}. Furthermore, three-body decays through effective five-dimensional operators including Higgses can be phenomenologically relevant in a variety of theories, from weak scale supersymmetry (e.g. \cite{Djouadi:1995gv,Barradas:1996xb}) to theories with exotic representations (e.g. \cite{Babu:2009aq,Ghosh:2018drw}), and motivate looking for new particles decaying to, for example, $H+q\bar q$ or $H+\ell\ell$. It is more than worth understanding all the channels in which searches for final states with Higgs resonances may improve our coverage.

Despite our hopes and expectations, new physics has proven hard to find at the LHC. We hit paydirt with the discovery of the Higgs, but our intellectual curiosity is far from sated and drives us to continue digging. Looking throughout the parameter space of expected signals remains useful, and the fantastic constraints already placed on many, many models are a testament to experimental efforts. We have here suggested a new search strategy for electroweak states which makes fundamental use of our most recent, exciting discovery. How elegant it would be for Higgs resonances to once more lead us to the next great treasure of particle physics.

\begin{acknowledgments} 
We are indebted to Nathaniel Craig for suggesting this study, and we thank him and Carlos Wagner for comments on a draft of this manuscript. This work is supported in part by the US Department of Energy under the awards DE-SC0014129 and DE-SC0011702. 
SK was supported in part by a Mafalda and Reinhard Oehme Postdoctoral Fellowship from the Enrico Fermi Institute at the University of Chicago. SK and U\"{O} are grateful for the support of a Worster Fellowship from the physics department at UCSB, and U\"{O} thanks the UCSB College of Creative Studies for the support of a SURF Fellowship.
\end{acknowledgments}

\appendix*
\section{CMS Multilepton Search Reinterpretation}\label{app:reinterp}
\begin{figure}[h]
        \centering
        \includegraphics[width = 8. cm]{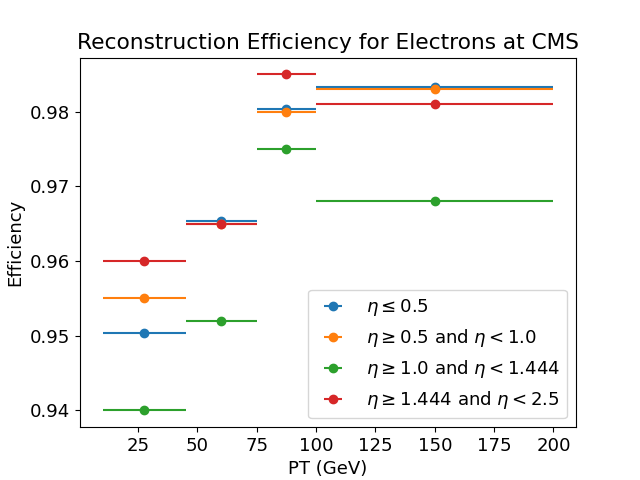}
            \caption{The Electron Identification Efficiency we use for the Delphes Card. \label{fig:eleff}}
\end{figure}

\begin{figure}[h]
        \centering
        \includegraphics[width = 8. cm]{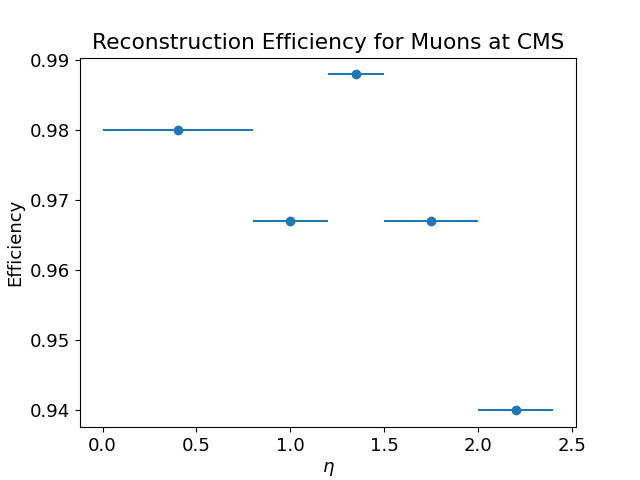}
            \caption{The Muon Identification Efficiency we use for the Delphes Card. \label{fig:mueff}}
\end{figure}

The most stringent constraints on these models from the LHC come from searches for multilepton final states. We reinterpeted the 13 TeV CMS search \cite{Sirunyan:2019bgz} with $137 \text{ fb}^{-1}$ of data using expert mode of \texttt{MadAnalysis 5} \cite{Conte_2013} to implement cuts and event selection. We simulated events in \texttt{MadGraph 5}, passed them to \texttt{Pythia} for decays and showering, and then to \texttt{Delphes} for detector simulation. The cuts and event selection of the type-III seesaw signal regions were implemented faithfully, except that the impact parameter requirements were ignored. We found that it was crucial to use the same PDF as the CMS analysis, namely NNPDF3.0 (LHAPDF ID: 261000) as described in \cite{Ball_2015}. Another important ingredient was the lepton reconstruction efficiency, for which we could not find exact specifications. We based our electron and muon efficiencies mainly on \cite{CMS-DP-2019-022} and \cite{Rembser:2019ijh} respectively, and implemented them as shown in Figures \ref{fig:eleff}, \ref{fig:mueff}. 

\begin{figure}[t]
        \centering
        \includegraphics[width = 7.8 cm]{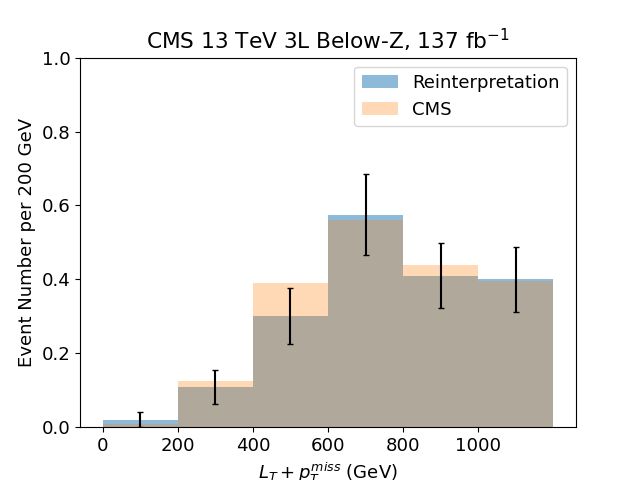}
    \caption{Comparison of the 3L below-Z signal region between the simulation performed in the CMS analysis \cite{Sirunyan:2019bgz} for the $M_\Sigma = 700 \text{ GeV}$ point and our simulation of the same model implemented in our analysis pipeline. The black intervals denote our $1\sigma$ statistical uncertainty bands.}           
    \label{fig:3LbZ}
    \end{figure}

\begin{figure}[t]
        \centering
        \includegraphics[width = 7.8 cm]{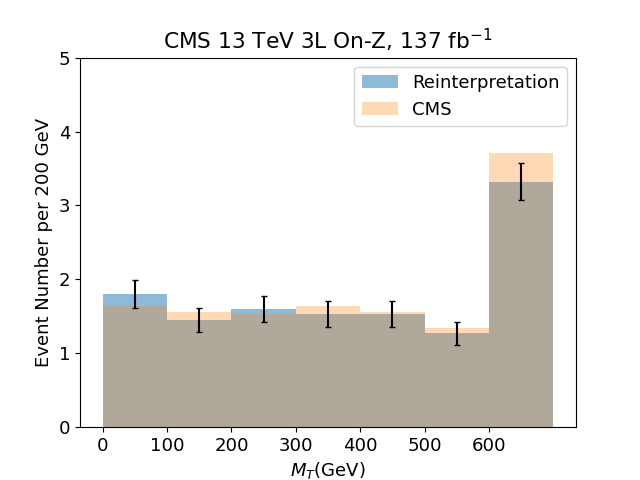}
     \caption{The same as Figure \ref{fig:3LbZ} for the 3L on-Z signal region.}   
    \label{fig:3LonZ}
    \end{figure}

\begin{figure}[t]
        \centering
        \includegraphics[width = 7.8 cm]{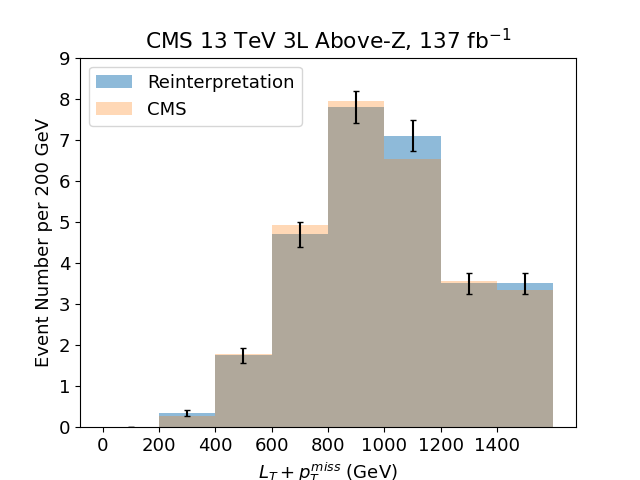}
     \caption{The same as Figure \ref{fig:3LbZ} for the 3L above-Z signal region.}       
    \label{fig:3LaZ}
    \end{figure}

\begin{figure}[t]
        \centering
        \includegraphics[width = 7.8 cm]{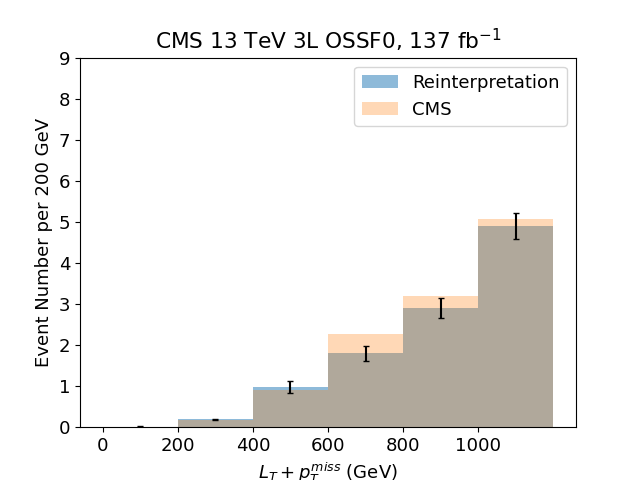}
    \caption{The same as Figure \ref{fig:3LbZ} for the 3L OSSF0 signal region.} 
    \label{fig:3Lossf0}
    \end{figure}

\begin{figure}[t]
        \centering
        \includegraphics[width = 7.8 cm]{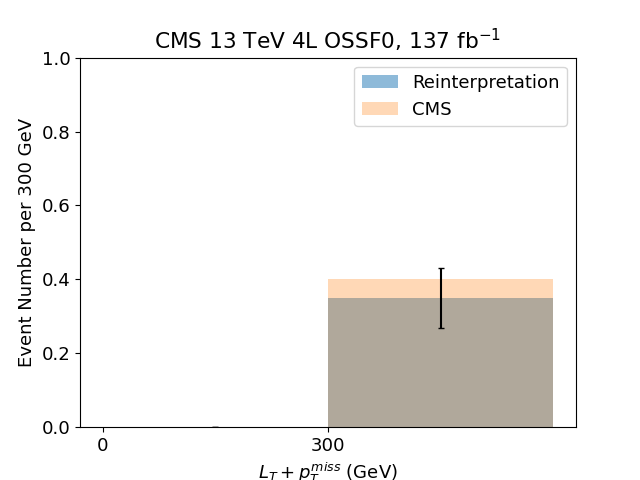}
    \caption{The same as Figure \ref{fig:3LbZ} for the 4L OSSF0 signal region.} 
    \label{fig:4Lossf0}
    \end{figure}

\begin{figure}[t]
        \centering
        \includegraphics[width = 7.8 cm]{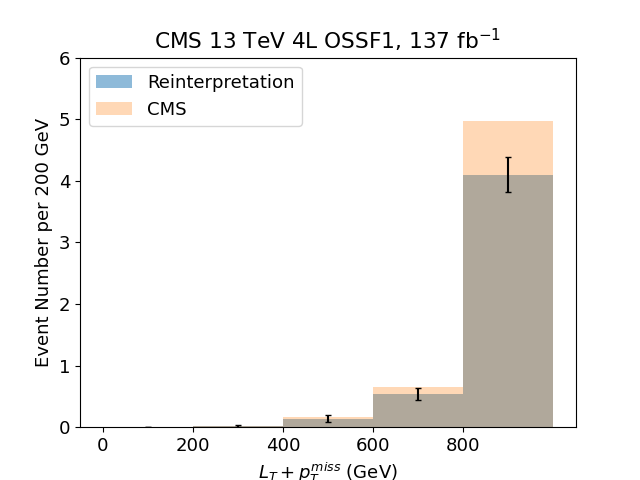}
     \caption{The same as Figure \ref{fig:3LbZ} for the 4L OSSF1 signal region.}     
    \label{fig:4Lossf1}
    \end{figure}

\begin{figure}[t]
        \centering
        \includegraphics[width = 7.8 cm]{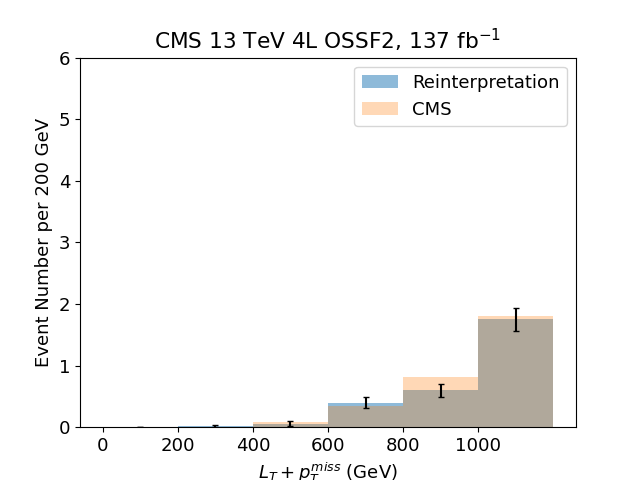}
    \caption{The same as Figure \ref{fig:3LbZ} for the 4L OSSF2 signal region.}    
    \label{fig:4Lossf2}
    \end{figure}

To validate our reinterpretation, we simulated 100k $p p \rightarrow \Sigma \bar \Sigma$ events in the Type-III flavor democratic seesaw model for which the CMS analysis \cite{Sirunyan:2019bgz} presents simulation results. In Figures \ref{fig:3LbZ}-\ref{fig:4Lossf2} we compare the results of our simulation and reinterpretation to those of CMS for the parameter point $M_\Sigma = 700 \text{ GeV}$. 

After validating our simulation and reinterpretation pipeline, we derive upper limits at 95\% confidence level on the number of signal events in any signal region assuming Poisson distributions for the signal and background, and ignoring the uncertainty in the backgrounds. We use data found in the HEPData repository entry for this search\footnote{https://www.hepdata.net/record/ins1764474}. Lacking the full covariance matrix for the many signal regions, we instead conservatively say that a model point is ruled out at 95\% confidence if it predicts more events than the 95\% upper limit in any individual bin. As expected, the bins with the highest values of the kinematic discriminants tend to give the most stringent upper limits. The 95\% confidence limits placed from this reinterpretation can be found in Table \ref{tab:reintbounds}.
\par
There are 7 exclusive signal regions of relevance for our reinterpretation of the CMS search. The events are first distinguished based on whether they contain 3 leptons (`3L') or at least 4 leptons (`4L'). The events with 3 leptons are further divided into four distinct signal regions: those that have an opposite sign same flavor (OSSF) lepton pair with invariant mass below the `Z mass window' $M_Z \pm 15  \text{ GeV}$ (`Below-Z'), those that have an OSSF pair within the Z mass window (`On-Z'), that have an OSSF pair above the Z mass window (`Above-Z'), and those with no OSSF pair at all (`OSSF0'). 
%For example an $L_T$ value of 400 in the graph for the 3L Below-Z region corresponds to the bin $200< L_T \leq 400 $.  
The events with 4 or more leptons are similarly divided into regions that have 0, 1, and 2 OSSF pairs. The signal regions are then binned based on a kinematic discriminant variable $L_T+p_T^\text{miss}$, which is the scalar sum of the $p_T$ of all charged final state leptons plus the missing energy. However for the 3L On-Z region, an alternate kinematic discriminant is found to increase sensitivity, so $M_T = \sqrt{2p^{miss}_T p^l_T (1-\cos(\Delta \phi_{p^{miss}_T,p^l_T}))}$ is used, where $p^l_T$ is the transverse momentum of the lepton which is not part of the OSSF pair.
    
%\bibliographystyle{plainurl}
\bibliography{higgsxbib}

\end{document}